\documentstyle[emulateapj,psfig]{article}
\def\solar{\ifmmode_{\mathord\odot}\else$_{\mathord\odot}$\fi}
\begin{document}

\voffset = -1.5truecm

\newcommand{\gray}{$\gamma$-ray\  } \newcommand{\grays}{$\gamma$-rays\
} \newcommand{\etal}{{\it  et al.\ }}  \newcommand{\lya}{Ly$\alpha$\ }
\newcommand{\epr}{e-print  astro-ph/  }  \newcommand{\e}  {\epsilon  }
\newcommand{\mic}{$\mu$m\ }

\title{Intergalactic Photon  Spectra from the  Far IR to the  UV Lyman
Limit for $0  < z < 6$ and  the Optical Depth of the  Universe to High
Energy Gamma-Rays}

\author{F.W.    Stecker}  \affil{NASA/Goddard  Space   Flight  Center}
\authoraddr{Greenbelt,     MD     20771;     Floyd.W.Stecker@nasa.gov}
\affil{Department of Physics  and Astronomy, University of California,
Los  Angeles}   \authoraddr{Los  Angeles,  CA90095-1547}  \author{M.A.
Malkan}  \affil{Department  of Physics  and  Astronomy, University  of
California,   Los  Angeles}  \authoraddr{Los   Angeles,  CA90095-1547;
malkan@astro.ucla.edu}
%\and 
\author{S.T.   Scully}  \affil{Department  of Physics,  James  Madison
University} \authoraddr{Harrisonburg, VA 22807; scullyst@jmu.edu}

\begin{abstract}

We calculate  the intergalactic photon  density as a function  of both
energy and redshift for  $0 < z < 6$ for photon  energies from .003 eV
to the Lyman  limit cutoff at 13.6 eV in  a $\Lambda$CDM universe with
$\Omega_{\Lambda} =  0.7$ and $\Omega_{m} = 0.3$.   The basic features
of our backwards evolution model for galaxies were developed in Malkan
and Stecker  (1998 and 2001).  With  a few improvements,  we find that
this evolutionary  model gives predictions  of new deep  number counts
from  {\it   Spitzer}  as  well  as  a   calculation  spectral  energy
distribution  of the  diffuse infrared  background which  are  in good
agreement  with the data.   We then  use our  calculated intergalactic
photon densities  to extend  previous work on  the absorption  of high
energy \grays  in intergalactic space  owing to interactions  with low
energy photons and the 2.7 K cosmic microwave background radiation. We
calculate the optical depth of the universe, $\tau$, for \grays having
energies from 4 GeV to 100 TeV emitted by sources at redshifts from ~0
to 5.   We also give an  analytic fit with  numerical coefficients for
approximating $\tau(E_{\gamma}, z)$.  As an example of the application
of our results,  we calculate the absorbed spectrum  of the blazar PKS
2155-304 at $z  = 0.117$ and compare it with  the spectrum observed by
the H.E.S.S. air Cherenkov \gray telescope array.

\end{abstract}

\section{Introduction}

The potential importance of  the photon-photon pair production process
in high energy  astrophysics was first pointed out  by Nikishov (1962)
before the discovery  of the CMB (2.7 K  cosmic microwave background).
However,  his   early  paper  overestimated  the   energy  density  of
intergalactic starlight  radiation by three orders  of magnitude. With
the discovery  of the CMB, Gould  \& Schreder (1966)  and Jelly (1966)
showed that the universe would be opaque to \grays of energy above 100
TeV at  extragalactic distances.  Stecker (1969) and  Fazio \& Stecker
(1970) included  cosmological  and   redshift  effects,  showing  that
photons  from a \gray  source at  a redshift  $z_{s}$ above  an energy
$\sim 100(1+z_{s})^{-2}$  TeV would be significantly  absorbed by pair
production interactions with the CMB.

Following the {\it CGRO}  (Compton Gamma-Ray Observatory) discovery of
the  strongly flaring  \gray blazar  3C279 at  redshift  0.54 (Hartman
\etal 1992),  and based on  earlier calculations by Stecker,  Puget \&
Fazio (1977) estimating the CIB (cosmic infrared background), Stecker,
de Jager  \& Salamon  (1992) proposed that  one can use  the predicted
pair  production  absorption  features  in blazars  to  determine  the
intensity of the  CIB, provided that the intrinsic  spectra of blazars
extend  to  TeV energies.   Subsequent  work  along  these lines  used
observations of  TeV spectra of blazars  to place upper  limits on the
CIB (Stecker \& de Jager 1993, 1997; Dwek \& Slavin 1994; Biller \etal
1998; Funk  \etal 1998; Vassiliev  2000; Stanev \&  Franceschini 1998;
Schroedter 2005).

With  the  advent  of  the  {\it COBE}  (Cosmic  Background  Explorer)
observations  (Dwek \& Arendt  1998; Fixsen  \etal 1997,  1998; Hauser
\etal 1998) real  data on the the CIB  became available.  Lower limits
from   galaxy  counts   help  in   determining  the   spectral  energy
distribution of  the CIB at wavelengths  where no {\it  COBE} data are
available.  In addition, the  upper limits from TeV \gray observations
and from  CIB fluctuations  help to determine  the CIB  at wavelengths
where no  direct measurements are  available (see Hauser \&  Dwek 2001
for  a review  of the  CIB measurements).   More  detailed theoretical
models of the  spectral energy distribution (SED) of  the CIB produced
by  cool stars  and  dust reradiation  in  galaxies were  subsequently
constructed (MacMinn \& Primack  1996; Franceschini \etal 1998; Malkan
\& Stecker 1998, 2001 (MS98,  MS01); Kneiske \etal 2002).  Such models
of  the  CIB can  be  used  to invert  the  original  approach and  to
calculate the expected  opacity of the universe to  high energy \grays
(Stecker  \etal 1992;  MacMinn \&  Primack 1996;  Stecker \&  de Jager
1998; Totani \& Takeuchi 2002;  Kneiske \etal 2004).  One can then use
these results  to derive the unabsorbed intrinsic  emission spectra of
TeV \gray sources.

The empirically based approach of  MS98 and MS01 yields CIB SEDs which
are  consistent  with  the  most  reliable  data  and  limits  on  the
CIB\footnote{There  was a  reported  large  CIB flux  at  100 \mic  by
Finkbeiner,  Davis   \&  Schlegel   (1999)  that  was   most  probably
contaminated by local solar system dust emission and is now considered
to  be an  upper limit  (Finkbeiner talk  at IAU  Symp. No.   204.)  A
reported near-IR  excess (Matsumoto  \etal 2005) conflicts  with upper
limits  from  TeV  blazar   spectra  (Schroedter  2005)  and  is  also
inconsistent with  galaxy counts and theoretical models  of early star
formation.   Other  theoretical problems  with  this reported  near-IR
excess are discussed by Madau  \& Silk (2005).  This near-IR radiation
may be  reflected sunlight from  interplanetary dust (Dwek,  Arendt \&
Krennrich 2005).  We do not  consider these reported fluxes to be part
of the CIB.},  although it should be noted that  these data have large
error  bars.  Malkan  \& Stecker  (2001)  also used  this approach  to
derive galaxy  LFs and source  counts at various  infrared wavelengths
which are found  to be in general agreement  with present data.  These
results have then been used to determine intrinsic (unabsorbed) source
spectra  of  well-observed  TeV  blazars  which can  be  shown  to  be
consistent with synchrotron self-Compton  emission models (De Jager \&
Stecker 2002;  Konopelko \etal  2003; Georganopoulos \&  Kazanas 2003;
Konopelko,  Mastichiadas \& Stecker  2005).  In  this way,  a synoptic
approach  to both  the \gray  and  infrared observations  can lead  to
knowledge  of both blazar  emission and  the low  energy intergalactic
photon background (Dwek \& Krennrich 2005).

It was  pointed out by  Madau \& Phinney  (1996) that the  optical and
ultraviolet radiation  produced by stars in galaxies  at redshifts out
to $\sim 2$ would make the  universe opaque to photons above an energy
of $\sim 30$  GeV emitted by sources at a redshift  of $\sim 2$, again
owing  to pair  production  interactions.  Salamon  \& Stecker  (1998)
(SS98) made detailed  calculations of the opacity of  \grays from such
interactions down  to energies of 10 GeV  and out to a  redshift of 3.
We continue and  expand this approach here using  recent data from the
{\it Spitzer} infrared observatory  and {\it Hubble} deep survey data.
We determine the IR-UV photon density  from 0.03 eV to the Lyman limit
at 13.6 eV for redshifts out  to 6 using very recent results from deep
surveys.   We will  refer to  this as  the  ``intergalactic background
light'' (IBL). We then use our results  on the density of the IBL as a
function  of redshift,  together  with the  opacity  of the  CMB as  a
function  of redshift,  to calculate  the opacity  of the  universe to
\grays for energies from 4 GeV to 100 TeV and for redshifts from $\sim
0$  to  5.  As  an  example  of the  application  of  our results,  we
calculate the  absorbed spectrum  of the blazar  PKS 2155-304 at  $z =
0.117$ and compare it with the spectrum observed by the {\it H.E.S.S.}
(High Energy Stereoscopic System) air Cherenkov \gray telescope array.
We  also give  an analytic  approximation to  our opacity  results for
$\tau(E_{\gamma},z)$.

\section{Calculation of the Intergalactic Background Light}

In order to calculate intergalactic IR photon fluxes and densities, we
again use  the same  method described in  detail by Malkan  \& Stecker
(MS98,  MS01).  This  method, a  "backwards evolution"  scheme,  is an
empirically based calculation  of the SED of the CIB  by using (1) the
luminosity  dependent  galaxy SEDs  based  on  observations of  normal
galaxy IR  SEDs by Spinoglio  \etal (1995), (2)  observationally based
luminosity  functions (LFs) staring  with the  work of  Saunders \etal
(1990) and  (3)  the latest  redshift  dependent luminosity  evolution
functions.\footnote{These  are empirically  derived curves  giving the
universal star formation rate (Madau \etal 1996) or luminosity density
(Lilly  \etal 1996)  as a  function  of redshift  which are  sometimes
referred  to  as  Lilly-Madau   plots.}   Since  MS98  and  MS01,  new
observations have  allowed us here  to make improvements to  our model
calculations as well as further detailed tests.

\subsection{Galaxy IR SEDs as a Function of Luminosity}

The  key  assumption  we make,  as  in  MS98  and  MS01, is  that  the
luminosity of a galaxy at all wavelengths, {\it viz.}, its SED, can be
predicted statistically  from its observed luminosity  in one infrared
waveband,  here  chosen  to  be  60$\mu$m.  This  is  founded  on  the
well-established fact that galaxies are  more luminous (now and in the
past)  when   they  have  higher  rates  of   recent  star  formation.
Empirically,  it  is  found  that  for  the  more  luminous  galaxies,
relatively more  of the energy from  these young stars  is absorbed by
dust grains  and re-radiated in the  thermal IR.  This  results in the
observational facts  that more luminous  galaxies have higher  IR flux
relative  to  optical  flux   and  warmer  IR  spectra.   These  clear
luminosity-dependent  trends  in  galaxy  SEDs  were  well  determined
locally from  the combination  of IRAS (Infrared  Astronomy Satellite)
and ground based  photometry for large ({\it e.g.},  all sky) samples.
Following our  previous work, we use the  quantitative measurements of
these  trends determined  by Spinoglio  \etal (1995)  and  Spinolgio ,
Andreani  and Malkan  (2002).  The  resulting  SEDs as  a function  of
galaxy luminosity  are based on broadband photometry  of IRAS selected
samples.  Since then, other  computations of IR backgrounds and source
counts have used  different sets of SEDs, based  on somehwat different
combinations  of data  and models,  in  some cases  estimated in  more
spectral  detail.  We have  therefore checked  whether these  new SEDs
might differ from  ours in either overall colors,  or in detail around
the  7--12$\mu$m region,  where  the strongest  spectral features  are
found.

The best example of these  new galaxy template SEDs has been published
by Xu  \etal (2001) in their  Table 2.  Many luminosity  bins could be
compared  with  our corresponding  SEDs  calculated  with the  scaling
formulae from Spinoglio \etal  (1995). For simplicity, we consider two
luminosity ranges near the knee of the galaxy luminosity function at $
z = 0$ and $ z = 1$.   So long as the agreement for SEDs near the knee
is  reasonable, the  final computed  IR backgrounds  will  also agree,
because they are dominated by galaxies around the knee.

\begin{figure*}[ht]
\epsfxsize=10cm \epsfbox{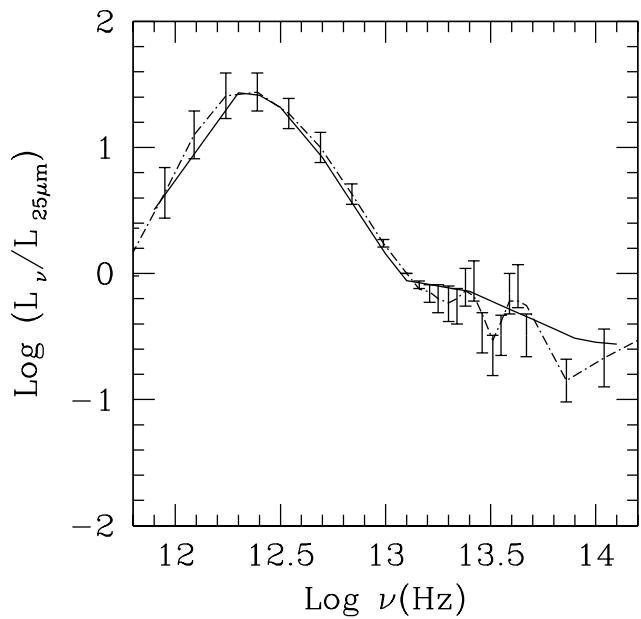}
\caption{SED, relative to flux at 25$\mu$m, for a galaxy with 25$\mu$m
luminosity of  $10^9 L\solar$. The  solid line shows our  model, while
the  error bars  show  the 1$\sigma$  range  of fluxes  from Xu  \etal
(2001),  which  are  also   shown  smoothed  to  broadband  photometry
resolution by the dotted line.}
\label{Xunrm9}
\end{figure*}

\begin{figure*}[ht]
\epsfxsize=10cm \epsfbox{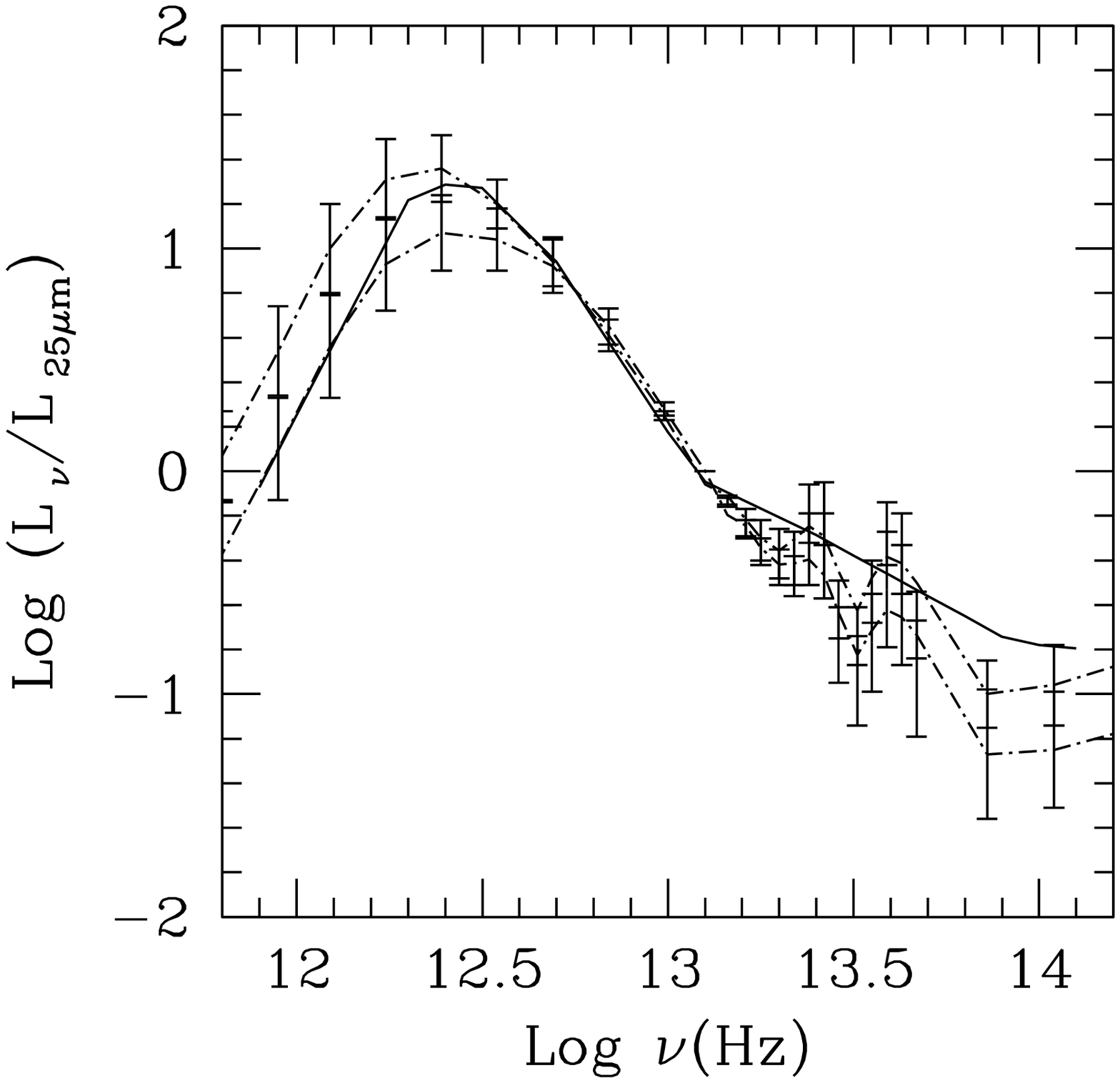}
\caption{SED, relative to flux at 25$\mu$m, for a galaxy with 25$\mu$m
luminosity of $10^{10} L\solar$. The solid line shows our model, while
the  error bars  show  the 1$\sigma$  range  of fluxes  from Xu  \etal
(2001),  which  are  also   shown  smoothed  to  broadband  photometry
resolution by  the dotted  lines.  The upper  set of points  shows the
``normal late-type  spiral" SED, while  the lower set of  points shows
the ``starburst galaxy" SED of the same 25$\mu$m luminosity, both sets
of points as given by Xu \etal (2001).}
\label{Xusbg10}
\end{figure*}

Figures \ref{Xunrm9} and \ref{Xusbg10} compare our computed SED (solid
line)  with the ranges  given by  Xu \etal  (shown by  1$\sigma$ error
bars) for  $10^9$ and $10^{10}$  solar luminosities at  25$\mu$m.  The
dashed line  shows the Xu  \etal SED smoothed  to the resolution  of a
typical  broad-band  filter, such  as  those  used  for {\it  Spitzer}
photometry.  All SEDs are normalized at 25$\mu$m.  For the latter more
luminous galaxies,  we plot the two  SEDs Xu \etal give  for their two
populations    of    galaxies,    normal   late-type    spirals    and
starbursts. Since  our simpler  computation includes only  one average
type of galaxy,  it could be expected to  straddle those two templates
from Xu \etal (2001).

The  agreeent  between  our SEDs  and  those  of  Xu \etal  (2001)  is
virtually perfect for all  wavelengths longward of 15$\mu$m.  The most
significant  disagreement  is  in   the  12$\mu$m  region,  where  our
broadband photometry  averages over the  two peaks at  higher spectral
resolution   produced  by   PAHs  (polycyclic   aromatic  hydrocarbon)
molecular  emission  bands (Peeters  \etal  2005)  and  the 10  $\mu$m
silicate  absorption feature (Chiar  \& Tielens  2006).  Even  in that
region,  the deviations  hardly ever  exceed the  $1\sigma$ dispersion
among Xu  \etal galaxy templates.   In any computation  which averages
over  redshifts,  such as  number  counts  and  especially in  diffuse
background calculations,  these SED  wiggles, with peak  amplitudes of
about 40\%, will make no observable difference in the results.

\subsection{The Local Infrared Luminosity Function}

\begin{figure*}[ht]
\epsfxsize=10cm \epsfbox{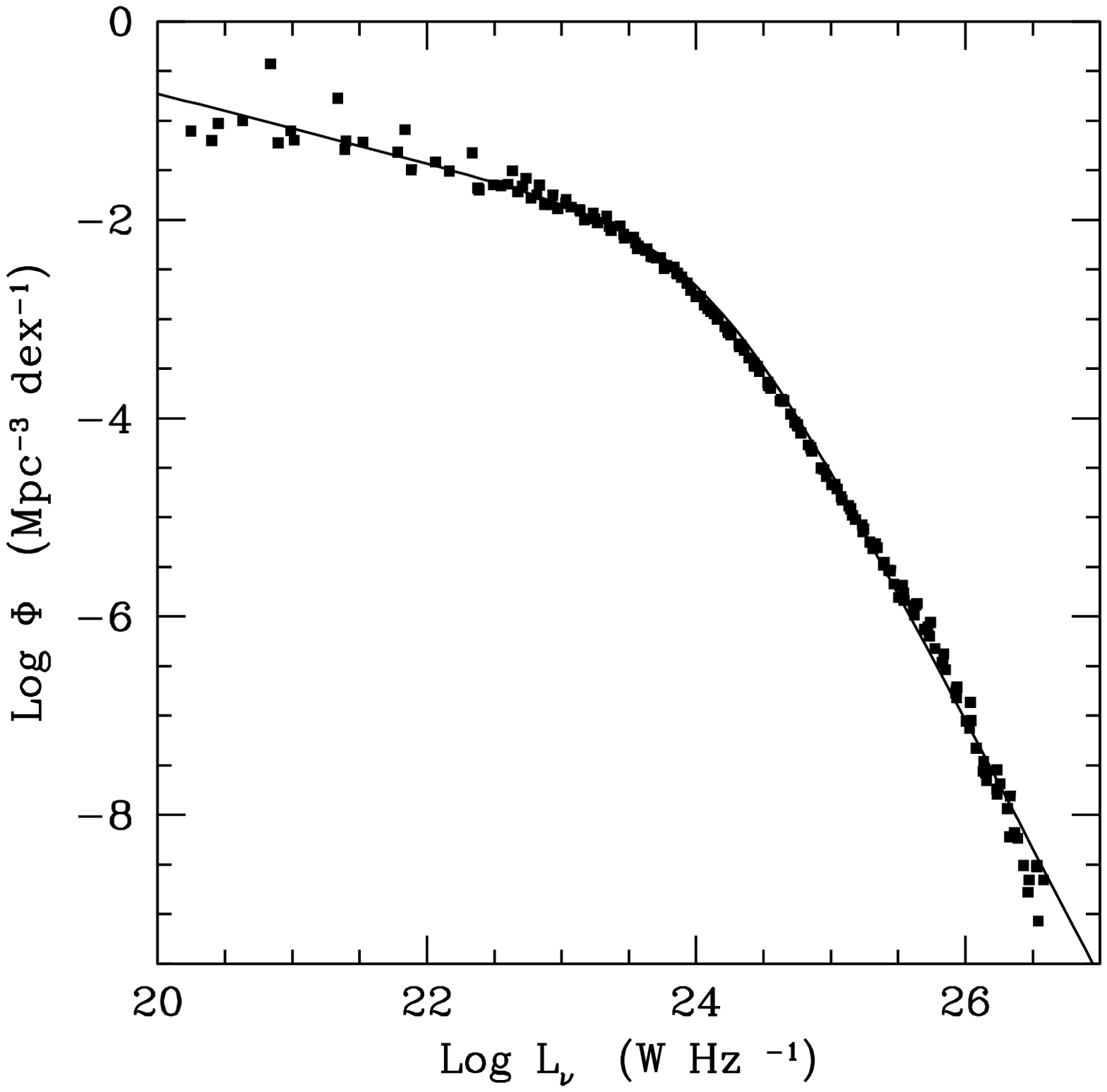}
\caption{60$\mu$m  local galaxy luminosity  function.  The  solid line
shows  our analytic fit,  as described  in the  text. The  square data
points show the compilation from Takeuchi \etal (2003).}
\label{LF}
\end{figure*}

The foundation of our backwards evolution calculation is an accurately
determined local infrared luminosity function of galaxies. As in MS01,
we  started  with the  local  luminosity function  at  60  \mic ,  the
wavelength where it  is best determined. But, instead  of the Lawrence
\etal (1986)  LF used in MS01,  we adopted the LF  from Saunders \etal
(1990), because it is based on  an extensive analysis of a larger data
sample. However,  we made an  update to the  Saunders LF based  on the
even more thorough  local IR LF determined by  Takeuchi, Yoshikawa and
Ishii (2003).  As  in MS98 and MS01, we took  an analytic function for
the local 60 \mic LF of the form

 $$\Phi (L,z=0)_{60} \ \propto x^{-a} (1 + {{x}\over{b}})^{-b} , \ \ x
\equiv L/L_{*}$$

\noindent which fits  the observational data at 60  \mic. LFs at other
wavelengths  were   obtained  by   using  the  average   template  SED
appropriate for  each luminosity  as is discussed  in MS01.   The only
difference in this paper was that we used an asymptotic low-luminosity
power-law  index of  $a  = 1.35$  in  the differential  LF which  then
steepens by  $b =  2.25$ at high  luminosities.  This LF  takes better
account of  the large  number of fainter  galaxies that are  now known
(Blanton \etal  2005 and  references therein.)  The  60 $\mu$m  LF was
normalized to  $8.9 \times 10^{-3}$ Mpc$^{-3}$ dex$^{-1}$  at the knee
luminosity $L_\star$  = $10^{23.93}$ W Hz$^{-1}$, as  derived from the
LF  of  Saunders  \etal (1990)  after  rescaling  to  $h =  0.7$.   In
addition, in this new calculation, we adopted a $\Lambda$CDM cosmology
with $\Omega_{\Lambda} = 0.7$ and $\Omega_{m} = 0.3$.  Figure \ref{LF}
compares our  adopted analytic luminosity function  for $ z  = 0$ with
the  data  summarized by  Takeuchi  \etal  (2003).   As expected,  the
agreement is perfect.

As in MS01,  the four luminosity relations obtained  by Spinoglio {\it
et al.} (1995) at 12, 25, 60 and 100 $\mu$m\ (and our estimates at 2.2
and 3.5 $\mu$m)  were inverted so that a luminosity  at any given rest
wavelength  could  be  determined   from  the  60  $\mu$m  luminosity,
$L_{60}$.  This allowed us to make a mapping of the 60 $\mu$m LF shown
in  Figure \ref{LF}  to an  LF at  any infrared  wavelength  using the
transformation relation

$$\Phi_{\lambda}(\log  L_{\lambda})  =  \Phi_{60}(\log L_{60})  (d\log
L_{60} / d\log L_{\lambda})$$

\noindent which MS01  showed were in good agreement  with local LFs at
other wavelengths.

\subsection{Evolution of the IBL SED with Redshift}

It is now well known that  galaxies had a brighter past owing to their
higher rates of  star formation and the fading  of stellar populations
as  they  age.   The   simplest  resulting  evolution  of  the  galaxy
luminosity function  is a  uniform shift in  either the  vertical axis
(number  density evolution),  or  in the  horizontal axis  (luminosity
evolution.)   For a  pure  power-law luminosity  function, number  and
luminosity  evolution  are  mathematically  equivalent.   In  reality,
however,  to  avoid unphysical  divergences  in  the  total number  or
luminosity of galaxies, the luminosity function must steepen at high L
and  flatten  at  low L.   Thus  real  LF's  will  have at  least  one
characteristic ``knee"  separating the  steep high-L portion  from the
flatter  low-L slope  (see  Figure \ref{LF}.   For  typical LF's  this
results in most of the  luminosity being emitted by galaxies within an
order of magnitude of this  knee.  Thus large uncertainties and errors
in the  LF far from this knee  will hardly change most  of the results
(e.g., number counts and integrated diffuse backgrounds).

Strong  luminosity evolution  of galaxies,  {\it i.e.},  a substantial
increase in the luminosity of this knee with redshift, is consistently
found by many  observations relating IR luminosity to  the much higher
star formation rate at $z \sim 1$ and to the recent determination that
most UV-selected  galaxies at  $z \sim 1$  are also  luminous infrared
galaxies (Burgarella \etal 2006).

In addition  to the evolution  of galaxy luminosity, some  increase in
galaxy number density is expected owing to the hierarchical clustering
predicted by  cold dark matter models.   However, luminosity evolution
is so strongly  dominant that it is difficult  to identify a component
of density evolution. MS98  and MS01 therefore assumed pure luminosity
evolution for their backwards evolution calculations.  Even five years
later,  this   is  still  an  excellent  description   of  the  latest
observations.  The  most recent galaxy luminosity  functions at higher
redshifts show  weak evidence for  a small number  evolution, combined
with very strong luminosity  evolution.  However, the number evolution
is still  sufficiently small (an increase  in space density  of only a
factor  of 1.5 to  1.9 times  the local  value), that  pure luminosity
evolution  (PLE) is  nearly as  good  a fit.   The PLE  model will  be
adopted here  because it  simplifies the calculation  and it  is quite
adequate or  the purposes of deriving intergalactic  low energy photon
densities  and spectra.   We base  our calculations  on  two plausible
cases of pure luminosity evolution  very similar to what MS98 and MS01
assumed:

\noindent (1)  In the more conservative scenario,  all galaxy 60$\mu$m
luminosities  evolved as  $(1+z)^{3.1}$. This  is the  same as  in the
"baseline  case" of  MS01 and  MS98 (with  their  luminosity evolution
parameter $ q = 3.1$), but  with evolution stopped at $z_{flat} = 1.4$
and galaxy  luminosities assumed constant (nonevolving)  at the higher
redshifts $1.4 <  z < 6$, with negligible  (assumed zero) emission for
$z  > 6$.   This  later assumption  is  supported by  the recent  {\it
Hubble} deep survey results (Bunker \etal 2004 and references therein;
Bouwens, R.J.,  Illingworth, G.D., Blakeslee, J.P. \&  Franx, M. 2005)
which indicate  that the average  star formation rate is  dropping off
significantly at a redshift of 6. Independent evidence from luminosity
functions of Lyman $\alpha$-emitting objects  at redshifts from 3 to 6
shows a similar decrease (Kashikawa \etal 2005).
\footnote{According to  Bouwens \etal (2005), the  star formation rate
at $z = 6$  is only about 70\% of that at $z =  1.5$.  We have run the
case where  this is  simulated by a  negative evolution in  the galaxy
number  density $\propto (1  + z)^{-0.637}$.   As expected,  this only
lowers our  results by $\sim  15$\%, an uncertainty which  is somewhat
counterbalanced by our assuming no star formation at redshifts greater
than 6.  This  overall uncertainty is less than  that arising from the
uncertainty in the stellar metallicity at these redshifts.}
 
\noindent (2)  The ``fast  evolution'' case where  galaxy luminosities
evolved as  $(1+z)^4$ for $0 < z  < 0.8$ and evolved  as $(1+z)^2$ for
$0.8 < z < 1.5$  with no evolution (all luminosities assumed constant)
for for $1.5 < z < 6$  and zero luminosity for $z > 6$. This evolution
model is based on  the mid-IR luminosity functions recently determined
out to $ z = 2$ by Perez-Gonzalez \etal (2005).

The  ``fast evolution''  picture is  favored by  recent  {\it Spitzer}
observations  (Le Floc'h  \etal 2005,  Perez-Gonzalez \etal  2005). It
provides a better description of  the deep {\it Spitzer} number counts
at 70 and 160$\mu$m than the ``baseline'' model.  However, {\it GALEX}
observations  indicate that the  evolution of  UV radiation  (see next
section) for  $0 < z < 1$  may be somewhat slower  and more consistent
with  the   ``baseline''  model  within   errors  (Schiminovich  \etal
2005). And the 24$\mu$m {\it  Spitzer} source counts are closer to the
baseline model  than the fast  evolution one.  The {\it  Spitzer IRAC}
(Infrared Array  Camera) counts lie  in between these two  models.  In
any case, we shall see that the possible small difference in evolution
has only a small effect on  the \gray opacity in the 10-100 GeV range,
as can be seen  from our results in Section 3.2. In  any case, we have
not  used a  simple evolution  model  for the  optical-UV, but  rather
adopted the results of SS98 (see next section). We also note that dust
extinction  effects at  the  higher redshifts  in  principal may  help
account for the differences between  the {\it GALEX} and {\it Spitzer}
results (see Section 4.2).

Fazio \etal (2004) have compared our source count predictions as given
in MS01 with the source count  data from {\it Spitzer} in the 3-8 \mic
range. Their paper shows that our model, although not giving a perfect
fit, fits these data as well as other models which have been proposed.
We  will present  our new  source  count calculations  (which are  not
substantially  different from  those given  in MS01)  for  the 24\mic,
70\mic and  160\mic {\it  Spitzer IRAC} bands  in a separate  IR paper
(Malkan \& Stecker, in preparation).

\subsection{The Intergalactic Optical and Ultraviolet Photon Spectra}

Throughout  the  substantial  evolution  of  galaxy  IR  luminosities,
backwards  evolution assumes  that the  same local  monotonic relation
between galaxy luminosity  and SED continues to hold.   In effect, the
relation between evolved stars  (and stellar mass) and their thermally
reradiated luminosity from dust grains  is assumed to remain the same.
The rest-frame optical and UV  continuum has a far larger contribution
from  more  massive  stars,  which  fade rapidly  over  cosmic  times.
Therefore stellar evolution is  expected to produce redshift evolution
in   the  optical-UV   SEDs  of   galaxies,  further   justifying  the
observationally-based  PLE models discussed  in the  previous section.
This  is reflected  in  the fact  that  our assumed  PLE models,  when
combined with the  library of galaxy SEDs as  a function of luminosity
(given explicitly  in MS98)  results in an  automatic increase  in the
ratio  of FIR  (dust) to  NIR  (stellar) emission  from galaxies  with
redshift.

\begin{figure*}[ht]
\epsfxsize=10cm \epsfbox{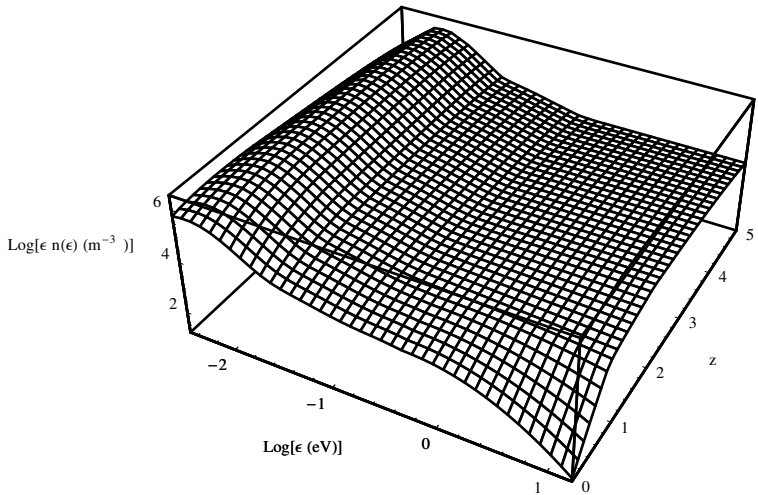}
\caption{The  photon   density  $\epsilon  n(\epsilon)$   shown  as  a
continuous function of energy and redshift.}
\label{3Dphot}
\end{figure*}

\begin{figure*}[ht]
\epsfxsize=10cm \epsfbox{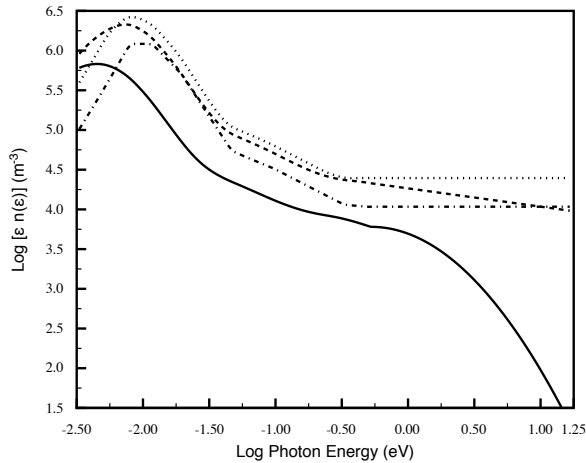}
\caption{The photon  density $\epsilon n(\epsilon)$ %in  $m^{-3}$ as a
function of energy  for various redshifts based on  the fast evolution
model for  IR evolution. The solid  line is for  $ z = 0$,  the dashed
line is for $ z = 1$ , the dotted line is for $ z = 3$, the dot-dashed
line is for $ z = 5$.}
\label{photz}
\end{figure*}

Our  calculation of  the diffuse  IR background  as  described earlier
extends  up   to  a  rest   frequency  of  $\log  \nu_{Hz}   =  14.1.$
corresponding to an  energy of $\sim 0.5$ eV. This  is the location of
the peak in the spectral  energy distributions of most galaxies and is
produced by  the light  of red giant  stars.  The spectrum  (in energy
density  units) then  curves downward  rapidly to  higher frequencies,
with  modest dependence  on galaxy  luminosity.  Although  galaxy SEDs
have a peak at  this energy, this peak, as opposed to  the far IR peak
in galaxy  SEDs, only manifests itself  as an inflection  point in the
photon density  spectrum.\footnote{We note that  the energy dependence
of the  differential photon energy  spectrum, $dn_{\gamma}/d\epsilon$,
is obtained  by dividing the  SED by the  square of the  photon energy
($\e^2$)  so that  the  starlight ``peak''  in  the SED  has very  few
photons compared to the dust reradiation peak in the far infrared.} At
wavelengths  shortward  of this  near  IR-optical  ``peak' the  photon
density spectra drop steeply (See Figures \ref{3Dphot} and \ref{photz}
which plot $\epsilon  n(\epsilon)$ for the case of  the fast evolution
model.)

To  estimate the  redshift-dependence of  stellar optical-UV  SEDs, we
employed an analytic approximation  to the more sophisticated near IR,
optical and UV SEDs used in SS98.  These SEDs were based on the Bruzal
\&  Charlot  (1993) stellar  population  synthesis  models for  galaxy
evolution.  The assumptions made by SS98 for SEDs at various redshifts
are now supported by various observational results on galaxy evolution
in the near IR-UV (Cowie \etal 1996, Pozzetti \etal 2003, Malkan, Webb
\& Konopacky 2003, Havens \etal  2004, Yee \etal 2005).  We adopt here
SEDs which  are roughly halfway  between the ``with''  and ``without''
metallicity  correction  cases  given  in SS98.   This  assumption  is
supported by the recent results of  Yan \etal (2005) who find that the
metallicity of stars in high redshift galaxies (out to $z = 6$) is not
extremely  low and  that  some heavy  element  production has  already
occurred by  this time.   In addition, a  recent detection of  the 158
$\mu$m CII  fine structure line  in a quasar  at $z = 6.4$  suggests a
significant stellar metallicity existed  at redshifts as high as $\sim
6$ (Maiolino \etal  2005). Yan \etal (2005, 2006)  found that galaxies
at  $z \simeq  6$ have  a similar  blue color  to those  shown  by the
spectral  evolution of the  population synthesis  models of  Bruzal \&
Charlot  (1993), with  some even  bluer than  the models  predict.  We
therefore  take the  galaxy SEDs  given for  $z =  3$ and  extend them
unchanged to  $z = 6$.   This will allow  us to calculate  the optical
depth of the universe to \grays out to a redshift of 5.

We normalized the  near infrared end of our  full-spectrum SEDs to the
near IR  parts of our  IR SEDs which  were calculated as  described in
Section 2.1.   For redshifts $0 < z  < 1$ we extended  them to optical
and ultraviolet using a  parabolic approximation to the energy density
SEDs $\log u_{\nu}$.  The SEDs used by SS98 in this redshift range can
be  approximated   by  a  function  whose   logarithmic  slope  $d\log
u_{\nu}/d\log \nu$ steepens from 0 at $\log \nu = 14.1$ to -2 at $\log
\nu$  =  15.0.   For $0  <  z  <  1$  one  can use  such  a  parabolic
approximation to the SED given by

$$ \log [ u_{\nu} / u_{\nu_{0}}] = \beta [\log (\nu /\nu_{0})]^2$$

\noindent where $\log \nu_{0} = 14.1$. The curvature parameter $\beta$
drops smoothly from -1.1 at $z = 0 $  to 0 at $z = 1$.  This is a good
approximation to  the more  sophisticated model calculations  of SS98,
adequate for  the purposes of  calculating photon densities  and \gray
optical depths.  At  higher redshifts we take account  of the shift in
galaxy colors  toward the blue  by making power-law  approximations to
the  galaxy  SEDs.   Denoting  the  power-law spectral  index  by  the
parameter $\alpha$, we take  $\alpha = -0.23(2 - z)$ for $  1 < z < 2$
and $\alpha = 0$ for $z > 2$.

These approximations to galaxy SEDs in the optical and UV reflect both
the stellar  population synthesis models (Bruzal \&  Charlot 1993) and
the direct Hubble space telescope observations that indicate that star
forming galaxies are bluer at $z > 0.7$ (De Mello \etal 2005).  At all
redshifts, because there  is a steep drop in galaxy  SEDs at the Lyman
limit of  13.6 eV  ($\log \nu  \simeq 15.5$) as  shown in  the spectra
given in SS98, we cut off our spectra entirely at this energy.  Such a
marked absence in photons at  wavelengths shortward of the Lyman limit
in  the spectra  of galaxies  at redshifts  $\sim 1$  was  observed by
Malkan, Webb \& Konopacky (2003).

\begin{figure*}[ht]
\epsfxsize=10cm \epsfbox{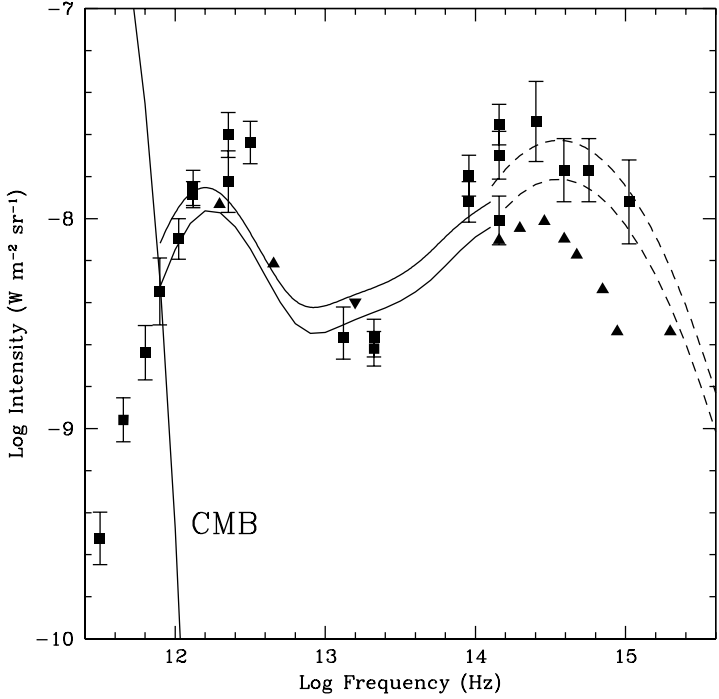}
\caption{Spectral  energy  distribution   of  the  diffuse  background
radiation at  $ z =  0$. Error bars  show data points,  triangles show
lower limits from number counts  and the inverted triangle is an upper
limit from  \gray observations (see  text). The upper and  lower solid
lines show our fast  evolution and baseline evolution predictions, and
the  dotted  lines  show  our  extensions  into  the  optical--UV,  as
described by SS98.  The steeply  dropping solid line near $10^{12}$ Hz
is the spectrum of the CMB.}
\label{IBL}
\end{figure*}

The IBL photon density, $\epsilon n(\epsilon) = \epsilon dn/d\epsilon$
for photon  energies from the  far IR to  the Lyman limit, given  as a
function of energy  and redshift is shown in  Figures \ref{3Dphot} and
\ref{photz} assuming the far-IR  fast evolution evolution model.  As a
check  on  the  two  models,  Figure  \ref{IBL}  shows  the  predicted
background SEDs from our model  calculations for $z = 0$ compared with
the  data and  empirical  limits. Most  of  the data  shown in  Figure
\ref{IBL}  can  be  found in  the  review  paper  by Hauser  and  Dwek
(2001). The inverted triangle shows the \gray upper limit from Stecker
and De Jager  (1997). The new lower limits in the  far-IR $\log \nu$ =
12.63 (70 \mic) and $\log \nu$  = 12.27 (160 \mic) are from Dole \etal
(2006).

\section{The Optical Depth of the Universe to Gamma-Rays}

\subsection{The Optical Depth from Interactions with CMB Microwave Photons}

The optical depth of the universe to the CMB is given by

%\begin{equation}
$$\tau_{CMB}  = 5.00 \times  10^5 \sqrt{{1.11  PeV}\over {E_{\gamma}}}
\int_0^z  {dz'~(1 +  z') ~{e^{-\left({1.11  PeV}\over  {E_{\gamma}(1 +
z')^2}\right)}}         \over         {\sqrt{\Omega_{\Lambda}        +
\Omega_{m}(1+z')^3}}}$$
%\end{equation}

\noindent for  the condition  $E_{\gamma} \ll 1.11/(1+z)^2$  PeV where
the interactions involve CMB photons on the Wien tail of the blackbody
spectrum. This  is an  update of the  formula given in  Stecker (1969)
using $T_{CMB}  = 2.73$  K and we  have taken a  $\Lambda$CDM universe
with $h = 0.7$. In all  of our calculations we use $\Omega_{\Lambda} =
0.7$ and $\Omega_{m} = 0.3$.

\subsection{The Optical Depth from Interactions with IBL Photons} 

\begin{figure*}[ht]
\epsfxsize=10cm \epsfbox{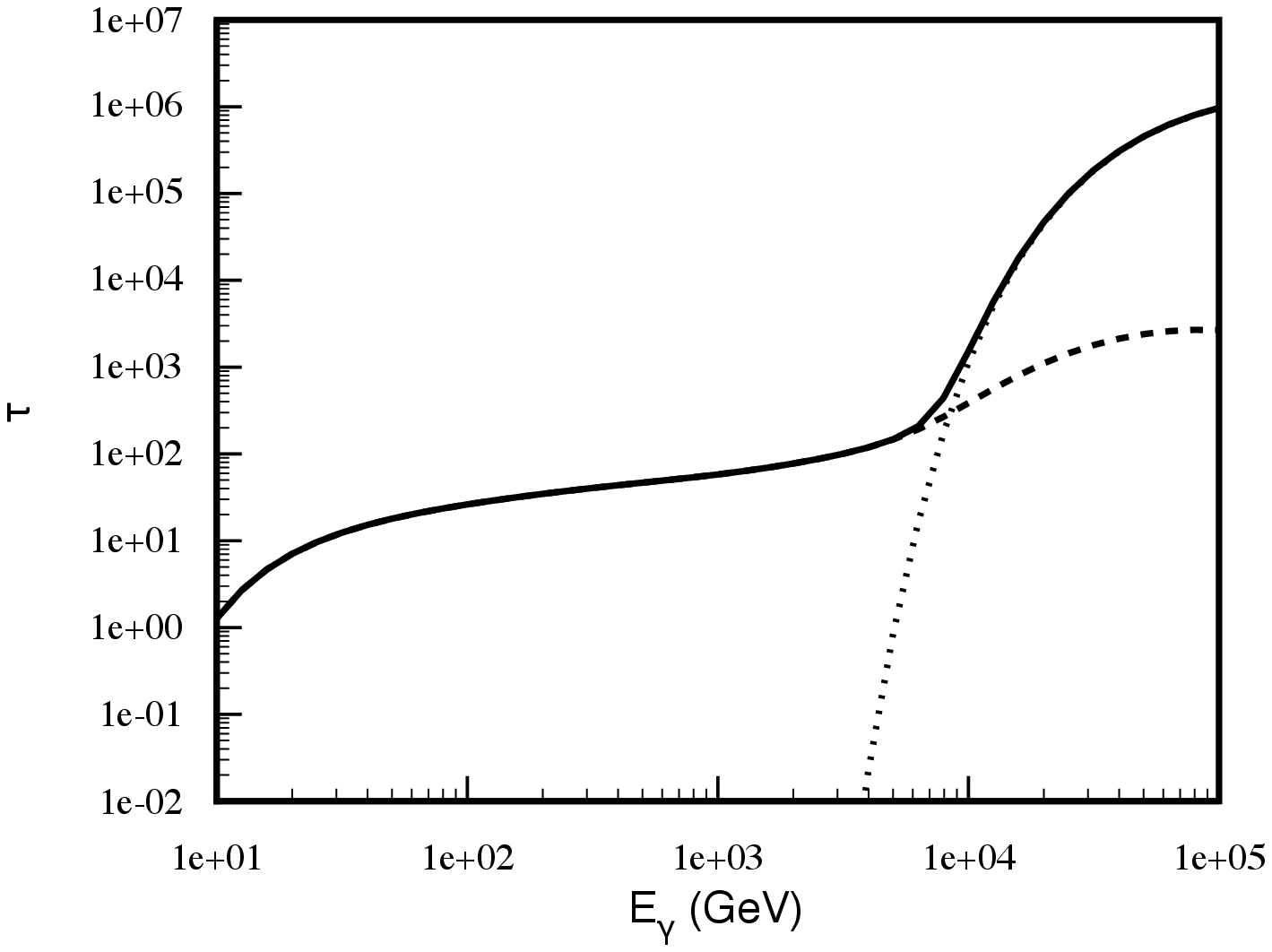}
\caption{The  optical  depth  of  the  universe  from  the  IBL  (fast
evolution case)  and the CMB as well  as the total optical  depth as a
function of energy for  a \gray\ source at a redshift of  3. It can be
seen that the contribution to the optical depth from the IBL dominates
at lower  \gray\ energies and that  from the CMB  photons dominates at
the higher  energies.  The  dashed curve is  for the  IBL contribution
alone and the dotted curve is for the CMB contribution alone.}
\label{z3}
\end{figure*}

To this result, we add the optical depth of the universe to the IBL as
a function  of $z$ as calculated  using the methods  described in SS98
and using our  IBL photon spectra as derived  in the previous section.
Figure \ref{z3} shows the  relative contributions to the optical depth
from the IBL and the CMB and the total optical depth for a source at a
redshift of 3.

\begin{figure*}[ht]
\epsfxsize=10cm \epsfbox{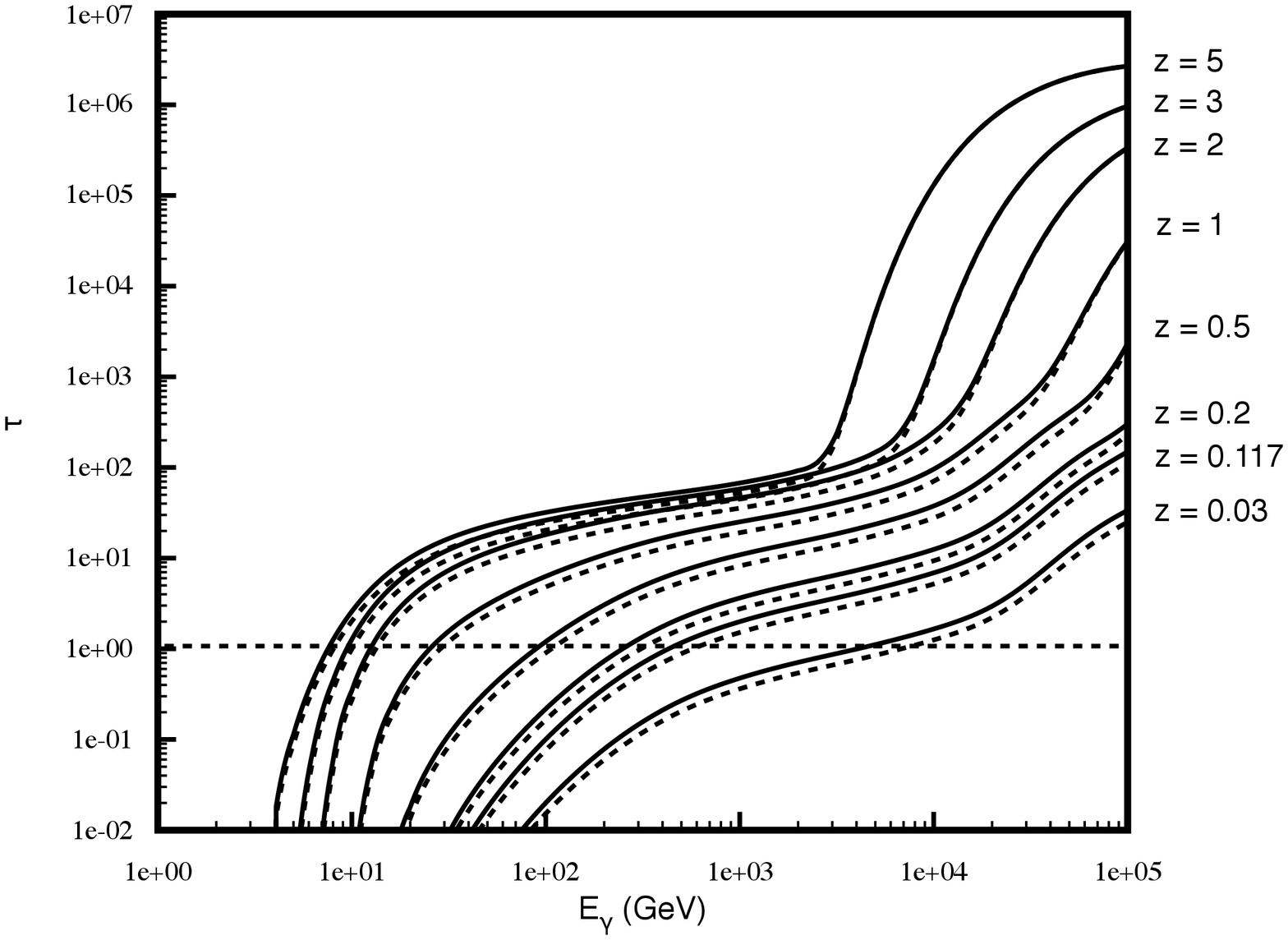}
\caption{The optical depth of the universe to \grays from interactions
with photons of  the IBL and CMB for \grays having  energies up to 100
TeV.   This is  given for  a family  of redshifts  from 0.03  to  5 as
indicated.  The solid  lines are for the fast  evolution IBL cases and
the dashed lines are for the baseline IBL cases.}
\label{taufam}
\end{figure*}

\begin{figure*}[ht]
\epsfxsize=10cm \epsfbox{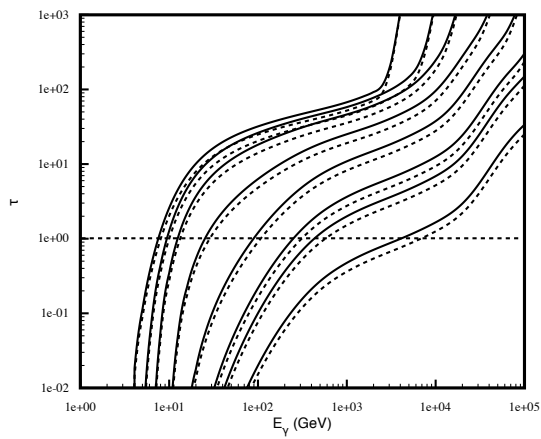}
\caption{The optical depth of  the universe to \grays from inteactions
with photons as in Figure \protect \ref{taufam} but truncated at $\tau
= 10^3$ to show detail.}
\label{trunctau}
\end{figure*}

Our results on  the optical depth as a function  of energy for various
redshifts out to a redshift of  5 are shown in Figure \ref{taufam} and
Figure \ref{trunctau}. Our new  results predict that the universe will
become  opaque  to \grays  for  sources  at  the higher  redshifts  at
somewhat  lower \gray  energies  than  those given  in  SS98. This  is
because the  newer deep surveys  have shown that there  is significant
star formation out to redshifts $z \ge 6$ (Bunker \etal 2004; Bouwens,
Illingworth,  Blakeslee and  Franx 2005),  greater than  the  value of
$z_{max} = 4$ assumed in SS98.

\begin{figure*}[ht]
\epsfxsize=10cm \epsfbox{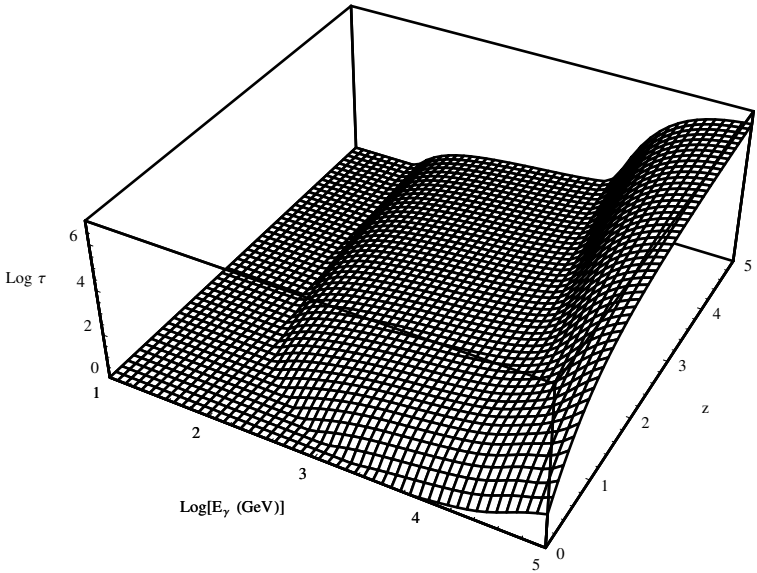}
\caption{The optical  depth of  the universe, $\tau  (E_{\gamma}, z)$,
given as  a continuous function of  \gray energy and  redshift for the
fast evolution IBL case with interactions with the CMB included.}
\label{3Dtau}
\end{figure*}

\begin{figure*}[ht]
\epsfxsize=10cm \epsfbox{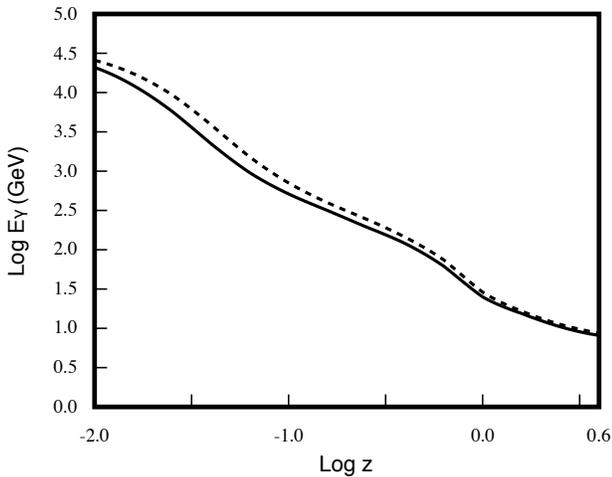}
\caption{The critical optical depth $\tau  = 1$ as a function of \gray
energy and redshift for the  fast evolution (solid curve) and baseline
(dashed curve)  IBL cases. Areas to  the right and  above these curves
correspond  to the  region where  the universe  is optically  thick to
\grays .}
\label{tau1}
\end{figure*}

Figure \ref{tau1} shows the energy-redshift relation giving an optical
depth $\tau = 1$ based  on our calculations of $\tau (E_{\gamma}, z)$.
This  curve is  generated by  the intersection  of the  function $\tau
(E_{\gamma}, z)$ shown in Figure \ref{3Dtau} with the plane defined by
the condition $\log \tau = 0$.  At energies and redshifts above and to
the  right  of   this  curve  the  universe  is   optically  thick  to
$\gamma$-rays. Similarly,  at energies and redshifts below  and to the
left  of  this  curve  the  universe is  optically  thin.   The  first
inflection point in the curve is  caused by the far-IR rollover in the
metagalactic  photon density as  shown in  Fig.  \ref{3Dphot}  and the
second  inflection point  is caused  by  the rollover  in the  optical
photon density, also as shown in Fig.  \ref{3Dphot}.

The function shown  in Figure \ref{tau1} is quite  different from that
produced by  Fazio \& Stecker  (1970) which was based  on interactions
with the  CMB alone.   This illustrates the  importance of the  IBL in
determining  the opacity  of the  universe  to high  energy \grays  at
higher redhsifts as  first pointed out by Stecker  \etal (1992).  This
figure  also illustrates  the extremely  large optical  depths  at the
higher energies due to interactions with the CMB.

We find that  the function $\tau (E_{\gamma}, z)$  can be approximated
by the analytic form

$$\log \tau = Ax^4+Bx^3+Cx^2+Dx+E $$

\noindent over  the range  $0.01 <  \tau < 100$  where $x  \equiv \log
E_{\gamma}$ (eV).  The  coefficients A through E are  given in Table 1
for  various redshifts.  This  analytic approximation  can be  used in
comparing our results with other work.

\begin{table*}[ht]
\centerline{Table 1: Coefficients for the Baseline Model Fits}
\begin{center}
\begin{tabular}{|cccccc|}  \hline \hline
$z$ &  $A$ &  $B$ &  $C$ &  $D$ & $E$  \\ \hline  0.03 &  -0.0151449 &
1.02602  & -24.2313  &  243.652 &  -893.883  \\ 0.112  & -0.0107295  &
-9679.05 & -138.498 & 3518.31 & -21256.9 \\ 0.2 & -0.0149538 & 1.02341
&  -24.2282 &  243.291 &  -888.586  \\ 0.5  & -0.0389542  & 2.12529  &
-42.9679 & 382.842 & -1270.59 \\  1 & -0.127954 & 6.22323 & -113.306 &
915.786 &  -2772.64 \\ 2 & -0.192839  & 9.13298 & -161.92  & 1274.05 &
-3753.66 \\ 3 & -0.143133 &  6.70614 & -117.706 & 917.64 & -2680.54 \\
5 & -0.281498 & 12.8979 & -221.364 & 1687.01 & -4816.33 \\

\hline
\end{tabular}
\end{center}
\end{table*}

\section{Implications of Results}

\subsection{The TeV Spectrum of PKS 2155-304}

Our  results for  $\tau (E_{\gamma},  z)$ can  be used  to  derive the
intrinsic \gray spectra of extragalactic sources. Based on the earlier
results of De  Jager \& Stecker (2002) for  low redshifts, the spectra
of the best observed extragalactic  TeV sources, {\it viz.} the BL Lac
objects  Mkn  501  and  Mkn  421, were  analysed  by  Konopelko  \etal
(2003). As an example here, we pick the next best observed source, the
blazar PKS 2155-304, which is situated  at a redshift of 0.117 and for
which there  are good  recent TeV spectral  data obtained by  the {\it
H.E.S.S.}  air Cerenkov \gray  telescope array (Aharonian \etal 2005).
This source  had earlier been reported  by the Durham group  to have a
flux above  0.3 TeV  of $\sim 4  \times 10^{-11}$  cm$^{-2}$ s$^{-1}$,
(Chadwick \etal 1999), close to that predicted by Stecker, de Jager \&
Salamon (1996) using a simple synchrotron self-Compton (SSC) model.

Using absorption results  obtained by Stecker and de  Jager (1998) and
assuming an approximate $dn/dE  \propto E^{-2}$ photon source spectrum
(corresponding to  a flat SED), which  would apply near  the very high
energy \gray Compton ``peak'' in  the quasi-parabolic log-log SED of a
typical SSC model (Stecker, de  Jager \& Salamon 1996), Stecker (1999)
predicted that this PKS 2155-304  would have its spectrum steepened by
$\sim$ 1 in its spectral index between $\sim 0.3$ and $\sim 3$ TeV and
would show a  pronounced absorption turnover above $\sim  6$ TeV.  The
recent {\it  H.E.S.S.}  observations  (Aharonian \etal 2005)  bear out
these predictions.

\begin{figure*}[ht]
\epsfxsize=10cm \epsfbox{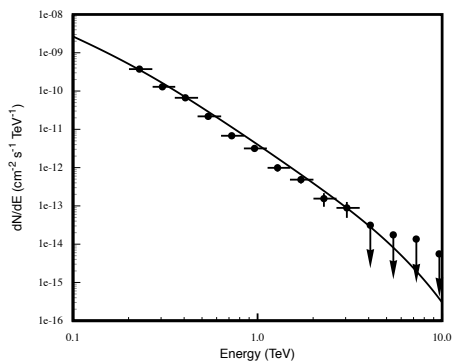}
\caption{The  \gray  data from  the  {\it  H.E.S.S.}  for  PKS2155-304
compared with  the theoretical spectrum for  PKS2155-304 calculated by
assuming an  unabsorbed source  spectrum proportional to  $E^{-2}$ and
multiplying  by   $e^{-\tau}$  using  $\tau(z=0.117)$   for  the  fast
evolution model  based on Perez-Gonzalez \etal (2005)  as discussed in
the text.}
\label{PKS2155}
\end{figure*} 
 
We revisit PKS 2155-304 here,  comparing the {\it H.E.S.S.}  data with
a photon  source spectrum  assumed to be  approximated by  an $E^{-2}$
which  was  then   steepened  by  $e^{-\tau(E_{\gamma})}$  with  $\tau
(E_{\gamma};  z =  0.117)$  as shown  in  Figure \ref{trunctau}.   The
resulting  spectrum is shown  in Figure  \ref{PKS2155} along  with the
{\it H.E.S.S.} data. We also found  that the best power law fit of our
calculated absorbed spectrum had a spectral index of -3.35 which is in
full  agreement  with  the  {\it  H.E.S.S.}  result  on  the  observed
spectrum of -3.32 $\pm$ 0.06 (Aharonian \etal 2005).  It can therefore
be seen that  the steep spectrum of PKS 2155-304  observed by the {\it
H.E.S.S.}   group  can  be  explained by  intergalactic  $\gamma\gamma
\rightarrow e^{+} e^{-}$ absorption.

We  further   note  that  a   steepening  in  an   $E^{-2}$  power-law
differential  photon  spectrum  by  one  power to  $E^{-3}$  would  be
produced   by   a   \gray    opacity   with   an   energy   dependence
$\tau(E_{\gamma}) \simeq \ln (aE_{\gamma})$.   In fact, we find such a
fit  over the energy  range 0.4  TeV$ <  E_{\gamma} <$  3 TeV  with an
approximate   relation  $\tau(E_{\gamma};  z   =  0.117)   \simeq  \ln
(7.3E_{\gamma})$ with $E_{\gamma})$ in TeV. An excellent fit over this
energy range at  $ z = 0.117 $ would  be $\tau(E_{\gamma}) \simeq [\ln
(5.52E_{\gamma})]^{1.26}$, which  is reflected in  a slight flattening
in our absorbed spectrum at the lower energies and a steepening at the
higher energies (see Figure \ref{PKS2155}).

\subsection{The Lyman Limit, UV Radiation at the Higher Redshifts,
 and Distant \gray Sources}

As previously stated, our SEDs in the UV range at the higher redshifts
are based on SS98 which is, in turn, based on the population synthesis
models of Bruzal \& Charlot  (1993). They give stellar SEDs at various
redshifts, but do  not take account of UV extinction  by dust. One way
of understanding  the somewhat smaller redshift evolution  of the star
formation rate  implied by the {\it  GALEX} observations (Schiminovich
\etal 2005)  {\it vs.} that  obtained from {\it  Spitzer} observations
(Le Floc'h \etal  2005, Perez-Gonzalez \etal 2005) is  that the effect
of dust extinction followed  by IR reradiation increases with redshift
(Burgarella, Buat \&  the {\it GALEX} team 2005).   Therefore, we may
have  overestimated the  UV photon  density at  the  higher redshifts.
This uncertainty should only  effect the absorption predictions in the
5 to 20 GeV energy range.

It  should be noted  that for  \gray sources  at the  higher redshifts
there is a  steeper energy dependence of $\tau  (E_{\gamma})$ near the
energy  where $\tau =  1$.  There  will thus  be a  sharper absorption
cutoff for sources at high redshifts.  It can easily be seen that this
effect is  caused by the  sharp drop in  the UV photon density  at the
Lyman limit, here approximated by an absolute cutoff.\footnote{Because
of the small photon flux above the Lyman limit, this approximation has
no significant effect on  our predicted \gray opacity values.}  Figure
\ref{3Dtau} shows the continuous function for the optical depth, $\tau
(E_{\gamma}, z)$.

\subsection{Implications for {\it GLAST} }

Because of the  energy dependence of absorption in  the blazar spectra
at the higher redshifts in the multi-GeV range, {\it GLAST}, the Gamma
Ray Large Space  Telescope ({\tt http://glast.gsfc.nasa.gov}; see also
McEnery, Moskalenko \&  Ormes 2004), will be able to  probe the IBL at
these redshifts and  probe the early star formation  rate (Chen, Reyes
\&  Ritz 2004).  For  example, {\it  GLAST} should  be able  to detect
blazars at $z \sim 2$ at multi-GeV energies and determine the critical
value  for  $E_{\gamma}$  above  which  absorption will  cut  off  the
spectrum, thereby distinguishing between  our predictions and those of
the various  models of Kneiske  \etal (2004).  More  importantly, such
{\it GLAST} observations  at redshifts $z \ge 2$  and $E_{\gamma} \sim
10$ GeV may complement the  deep galaxy surveys and help determine the
redshift when significant star formation began. Future {\it GLAST} 
observations in the 5 to 20 GeV energy range may also help to clarify 
the uncertainty in the amount of dust extinction pointed out 
in the previous section by determining the mean dsnsity of UV photons 
at the higher redshifts through their absorption effect on the
\gray spectra of high redshift sources.

In fact,  {\it GLAST} need  not have to  detect \gray sources  at high
redshifts in  order to aquire  information about the evolution  of the
IBL.   If the diffuse  \gray background  radiation is  from unresolved
blazars  (Stecker  \&  Salamon   1996),  a  hypothesis  which  can  be
independently  tested by {\it  GLAST} (Stecker  \& Salamon  1999), the
effects of IBL absorption will  steepen the spectrum of this radiation
at \gray energies above $\sim 10$  GeV (SS98), This is a direct result
of the  energy and redshift dependences of  the absorption coefficient
as clearly indicated in Figure \ref{tau1}.

\subsection{Implications for Fundamental Physics}

Stecker \& Glashow (2001) used observations of the Mkn 501 spectrum to
place constraints on violations  of Lorentz invariance. They concluded
that the evidence  for \gray absorption in the spectrum  of Mkn 501 at
an  energy  of 20  TeV  from  $\gamma\gamma  \rightarrow e^{+}  e^{-}$
interactions  placed  a  strong  constraint on  violation  of  Lorentz
invariance, {\it  viz.} a constraint  of about one part  in $10^{15}$.
Observing   extragalactic  \gray  sources   at  higher   energies  and
redshifts,   where  significant   attenuation   from  $\gamma   \gamma
\rightarrow e^{+} e^{-} $ interactions is expected, can provide a more
sensitive test  of Lorentz invariance  violation. This, in  turn, will
place constraints  on some quantum gravity and  extra dimension models
(Stecker 2003; Jacobson, Liberati, Mattingly, \& Stecker 2004).

\section{Comparison with Previous Work}

Our baseline case is quite  similar to the optical depths predicted by
Totani \& Takeuchi (2002) at  low redshifts, except that their assumed
galaxy SEDs had relatively less  mid-IR to optical emission than ours.
Our  flatter SEDs  result  in relatively  flatter $\tau  (E_{\gamma})$
curves than Totani \& Takeuchi.  However, at higher redshifts ($z = 1$
and 2), our optical depths  are substantially lower than the Totani \&
Takeuchi models, since  they assumed a very large  component of far-IR
emission  from elliptical galaxies  at $  z >3$.   Above `10  TeV, our
total  optical  depth  includes  the  contribution from  the  CMB  not
included by, Totani \& Takeuchi and is therefore much higher.

The  favored  ``best-fit'' model  of  Kneiske  \etal  (2004) gives  an
optical depth $\tau = 1$ at $E_{\gamma} \simeq 5 $ TeV for $z = 0.03$,
the redshift  of the  blazars Mrk 421  and Mrk  501.  This is  in good
agreement with our result. The form of the function $\tau (E_{\gamma};
z = 0.03)$ obtained by Kneiske  \etal (2004) is also in good agreement
with our result.  The results obtained  for $z = 0.2$ both here and by
Kneiske  \etal are also  in good  agreement. Our  results are  also in
accord with the model constraints given by Dwek \& Krennrich (2005).

On  the other  hand, Primack,  Bullock  \& Sommerville  (2005) find  a
consistently smaller optical  depth such that they find  $\tau = 1$ at
$E_{\gamma} \simeq  17$ TeV for $z =  0.03.$ At a redshift  of 0.2, we
find $\tau  = 1$  at $E_{\gamma}  \simeq 0.2 $  TeV and  Kneiske \etal
(2004) find $\tau = 1$ at  $E_{\gamma} \simeq 0.3 $ TeV. Primack \etal
(2005) find  $\tau  =  1$  at  $E_{\gamma}  \simeq  1$  TeV  for  this
redshift. The reason the  Primack \etal give consistently lower values
for $\tau(E_{\gamma})$ is because of  the lower flux which they obtain
for their model IR-SED. In this  regard, we point out that their model
flux at 15 $\mu$m is approximately  a factor of 2 lower than the lower
limit  obtained from  galaxy count  observations with  ISOCAM (Altieri
\etal 1999). In support of our higher opacity values, we note that the
spectrum of PKS2155-304 at $z  = 0.117$ shows our predicted steepening
from absorption at energies down to $\sim 0.3$ TeV where Primack \etal
(2005) would predict no significant absorption.
 
At  higher redshifts where  there is  more uncertainty,  Kneiske \etal
predict less absorption for  their ``best-fit'' model. For example, at
$z = 2$, their best fit model gives $\tau = 1$ at $E_{\gamma} \sim 50$
GeV, whereas we find $\tau = 1$ at $E_{\gamma} \sim 15$ GeV. It should
be noted, however,  that they consider a range of  models with a large
uncertainty in their absorption predictions at the higher redshifts.
Our calculations give a higher opacity at high redshifts because we
have taken account of the recent observational evidence for significant 
star formation out to a redshift of $\sim 6$ (Bunker \etal 2004) as
discussed in Section 2.3.

\section*{Acknowledgments} 

We wish to thank Pablo Perez-Gonzalez and Casey Papovich for helpful 
discussions of the {\it Spitzer} data.


\begin{thebibliography}{}

\bibitem[]{} Aharonian, F. \etal 2005, A\&A 430, 865

\bibitem[]{} Altieri, B. \etal 1999, A\&A 343, L65

\bibitem[]{} Biller \etal 1998, Phys. Rev. Letters 80, 2992

\bibitem[]{} Blanton, M.R. \etal 2005, ApJ 631, 208

\bibitem[]{} Bouwens, R.J., Illingworth, G.D., Blakeslee, J.P. \& 
Franx, M. 2005, ApJ, in press, \epr 0509641

\bibitem[]{} Bouwens, R.J.\& Illingworth, G.D. 2006, New Astron. Rev. 50, 152

\bibitem[]{} Bruzal, A.G. \& Charlot, S. 1993, ApJ 405, 538 

\bibitem[]{} Bunker, A.J., Stanway, E.R., Ellis, R.S. \& McMahon, R.G.
2004, MNRAS 355, 374

\bibitem[]{} Burgarella, Buat \&  the {\it GALEX} team 2005, in 
{\it The Dusty and Molecular Universe} (ESA SP-577)

\bibitem[]{} Burgarella, D. \etal 2006, A\&A 450, 69

\bibitem[]{} Chadwick, P.M. \etal 1999, Astropart. Phys. 11, 145

\bibitem[]{} Chen, A., Reyes, L.C. \& Ritz, S. 2004, ApJ 608, 686

\bibitem[]{} Chiar, J.E. \& Tielens, A.G.G.M. 2006, ApJ 637, 774

\bibitem[]{} Cowie, L.L., Songaila, A., Hu, E.M. \& Cohen, J.G.
1996, AJ 112, 839

\bibitem[]{} De Jager, O.C. \& Stecker, F.W. 2002, ApJ 566, 738

\bibitem[]{} De Mello, D.F., Wadadekar, Y., Dahlen, T., Casertano, S.
\& Gardener, J.P. 2005, Astron. J., in press, \epr 0510145

\bibitem[]{} Dole, H. \etal 2006, \epr 0603208

\bibitem[]{} Dwek, E. \& Arendt, R.G. 1998, ApJ 508, L9

\bibitem[]{} Dwek, E., Arendt, R.G. \& Krennrich, F. 2005, ApJ 635, 784

\bibitem[]{} Dwek, E. \& Krennrich, F. 2005, ApJ 618, 657

\bibitem[]{} Dwek, E. \& Slavin, J. 1994, ApJ 436, 696

\bibitem[]{} Fazio, G.G. \& Stecker, F.W. 1970, Nature 226, 135

\bibitem[]{} Fazio \etal 2004, ApJ Supp. 154, 39

\bibitem[]{} Finkbeiner, D.P., Davis, M. \& Schlegel, D.J. 1999,
ApJ 524, 867 

\bibitem[]{} Fixsen, {\it \etal.} 1997, ApJ 490, 482

\bibitem[]{} Fixsen, D.J., Dwek, E., Mather, J.C. 1998, 
Bennett, C.L. \& Shafer, R.A., ApJ, 508, 123

\bibitem[]{} Funk, N.M. \etal 1998, Astropart. Phys. 9, 97

\bibitem[]{} Georgonapoulos, M. \& Kazanas, D. 2003, ApJ 594, L27

\bibitem[]{} Gould, R.J. \& Schr\'eder, G. 1966, Phys. Rev. Letters 16, 252

\bibitem[]{} Hauser, M.G., {\it \etal.}, 1998, ApJ, 508, 25

\bibitem[]{} Hauser, M.G. \& Dwek, E., 2001, ARA\&A, 39, 249

\bibitem[]{} Havens, A. Panter, B., Jiminez, R. \& Dunlop, J. 2004,
Nature 428, 625

\bibitem[]{} Jacobson, T., Liberati, S., Mattingly, D. \& Stecker, F.W.
2004, Phys. Rev. Letters 93, 021101

\bibitem[]{} Jelly, J.V. 1966, Phys. Rev. Letters 16, 479

\bibitem[]{} Kashikawa, N. {\it \etal } 2005, submitted to ApJ

\bibitem[]{} Kneiske, T.M., Mannheim, K. \& Hartmann, D.H. 2002, A\&A 386, 1

\bibitem[]{} Kneiske, T.M., Bretz, T., Mannheim, K. \& Hartmann, D.H. 2004, 
A\&A 413, 807 

\bibitem[]{} Konopelko, A.K., Mastichiadis, A., Kirk, J.G., De Jager, O.C.
\& Stecker, F.W. 2003, ApJ 597, 851

\bibitem[]{} Konopelko, A., Mastichiadas, A. \& Stecker, F.W. 2005,
{\it Proc. 29th Intl. Cosmic Ray Conf. Pune, India}, p. 101,
\epr 0507479

\bibitem[]{} Lawrence, A. \etal 1986, MNRAS 219, 687

\bibitem[]{} Le Floc'h, E.\etal 2005, ApJ 632, 169

\bibitem[]{} Lilly, S.J., Le F\`evre, O., Hammer, F. \& Crampton, D.
1996, ApJ 460, L1

\bibitem[]{} MacMinn, D. \& Primack, J.R. 1996, Space Sci. Rev. 75, 413

\bibitem[]{} Madau, P. \etal 1996, MNRAS 283, 1388

\bibitem[]{} Madau, P. \& Phinney, E.S. 1996, ApJ 456, 124

\bibitem[]{} Madau, P. \& Silk, J. 2005, MNRAS 359, L37

\bibitem[]{} Maiolino, R. \etal 2005, A\&A, in press, \epr 0508064

\bibitem[]{} Malkan, M.A. \& Stecker, F.W. 1998, ApJ 496, 13

\bibitem[]{} Malkan, M.A. \& Stecker, F.W. 2001, ApJ 555, 641  

\bibitem[]{} Malkan, M.A., Webb, W. \& Konopacky, Q. 2003, ApJ 598, 878

\bibitem[]{} Matsumoto, T. \etal 2005, ApJ 626, 31

\bibitem[]{} McEnery, J.E., Moskalenko, I.V. \& Ormes, J.F. 2004,
in {\it Cosmic Gamma Ray Sources}, ed. K.S. Cheng \& G.E. Romero
(Dordrecht: Kluewer) p. 361, \epr 0406250

\bibitem[]{} Nikishov, A.I. 1962, Sov. Phys. JETP 14, 393

\bibitem[]{} Peeters, E, Mattioda, A.L., Hudgins, D.M. \& Allamandola, L.J.
2005, ApJ 617, L65  

\bibitem[]{} P\'erez-Gonz\'alez, P.G. \etal 2005, ApJ 630, 82

\bibitem[]{} Primack, J.R., Bullock, J.S. \& Sommerville, R.S. 2005,
\epr 0502177

\bibitem[]{} Salamon, M.H. \& Stecker, F.W. 1998, ApJ 493, 547

\bibitem[]{} Saunders, W. \etal 1990, MNRAS 242, 318

\bibitem[]{} Schroedter, M. 2005, ApJ, 628, 617

\bibitem[]{} Schiminovich, D. \etal 2005, ApJ 619, L47 

\bibitem[]{} Spinoglio, L., Malkan, M.A., Rush, B. Carrasco, L. \&
Recillas-Cruz, E. 1995, ApJ 453, 616 

\bibitem[]{} Spinoglio, L., Andreani, P. \& Malkan, M.A. 2002, ApJ 572, 105 


\bibitem[]{} Stanev, T., \& Franceschini, A., ApJ  494, L159

\bibitem[]{} Stecker, F.W. 1969, ApJ 157, 507

\bibitem[]{} Stecker, F.W. 1999, Astropart. Phys. 11, 83

\bibitem[]{} Stecker, F.W. 2001,  in
{\it The Extragalactic Infrared Background and its Cosmological 
Implications, IAU Symposium no. 204} ed. M. Harwit and M.G. Hauser,
p. 135

\bibitem[]{} Stecker, F.W. 2003, Astrpart. Phys. 20, 85

\bibitem[]{} Stecker, F.W. \& de Jager, O.C. 1993, ApJ 415, L71

\bibitem[]{} Stecker, F.W. \& de Jager, O.C. 1997,
in {\it Towards a Major Atmospheric Cerenkov Detector V, Proc.
Kruger Natl. Park Workshop on TeV Gamma-Ray Astrophysics} ed. O.C. de Jager
(Potchefstoom: Wesprint), e-print astro-ph/9710145

\bibitem[]{} Stecker, F.W. \& de Jager, O.C. 1998 
A\&A, 334, L85

\bibitem[]{} Stecker, F.W. 
de Jager, O.C. \& Salamon, M.H. 1992, ApJ, 390, L49

\bibitem[]{} Stecker, F.W. 
de Jager, O.C. \& Salamon, M.H. 1996, ApJ, 473, L75

\bibitem[]{} Stecker, F.W. \& Glashow, S.L. 2001, Astropart. Phys.
16, 97

\bibitem[]{} Stecker, F.W., Puget, J.-L. \& Fazio, G.G. 1977, ApJ 214, L51

\bibitem[]{} Stecker, F.W.  \& Salamon, M.H. 1996, ApJ, 464, 600

\bibitem[]{} Stecker, F.W.  \& Salamon, M.H. 1999, in 
{\it Proc. Intl. Cosmic Ray Conf., Salt Lake City}, ed. D. Kieda, M. 
Salamon \& B. Dingus, 3, 313, \epr 9909157

\bibitem[]{} Takeuchi, T.T., Yoshikawa, K. and Ishii, T.T. 2003,
ApJ 587, L89

\bibitem[]{} Totani,T.\& Takeuchi, T.T. 2002, ApJ 570, 470

\bibitem[]{} Vassiliev, V.V. 2000, Astropart. Phys. 12, 217

\bibitem[]{} Xu, C.K. \etal 2001, ApJ 562, 179

\bibitem[]{} Yan, H. \etal 2005, ApJ 634, 109

\bibitem[]{} Yan, H. \etal 2006, New Astron. Rev. 50, 127

\bibitem[]{} Yee, H.K.C., Hsieh, B.C., Lin, H. \& Gladders, M.D. 2005,
ApJ 629, L77

\end{thebibliography}
\end{document}